\documentclass[%
 article,
 amsmath,amssymb,
 aps,
onecolumn
noeqsecnum
]{revtex4-2}

\usepackage{graphicx}
\usepackage{dcolumn}
\usepackage{bm}
\usepackage{physics}

\begin{document}


\title{Thermal fluctuations and carrier localization induced by dynamic disorder in MAPbI\(_3\) described by a first-principles based tight-binding model}

\author{David J. Abramovitch}
 \affiliation{Department of Physics, University of California, Berkeley}
 \author{Wissam A. Saidi}
 \affiliation{Department of Mechanical Engineering and Materials Science, University of Pittsburgh}
\author{Liang Z. Tan}
\affiliation{Molecular Foundry, Lawrence Berkeley National Laboratory}


\begin{abstract}

Halide perovskites are strongly influenced by large amplitude anharmonic lattice fluctuations at room temperature. We develop a tight binding model for dynamically disordered MAPbI$_3$ based on density functional theory (DFT) calculations to calculate electronic structure for finite temperature crystal structures at the length scale of thermal disorder and carrier localization. The model predicts individual Hamiltonian matrix elements and band structures with high accuracy, owing to the inclusion of additional matrix elements and descriptors for non-Coulombic interactions. We apply this model to electronic structure at length and time scales inaccessible to first principles methods, finding an increase in band gap, carrier mass, and the sub-picosecond fluctuations in these quantities with increasing temperature as well as the onset of carrier localization in large supercells induced by thermal disorder at 300 K. We identify the length scale $L^*= 5$ nm as the onset of localization in the electronic structure, associated with associated with decreasing band edge fluctuations, increasing carrier mass, and Rashba splitting approaching zero.

\end{abstract}

\flushbottom
\maketitle

%
\thispagestyle{empty}

\section{Introduction}
Halide perovskites present an unusual combination of electronic and atomic structures \cite{Egger2018,Miyata2017, chu_low-frequency_2020}. The electronic structure in halide perovskites displays properties of a crystalline semiconductor, including ballistic charge transport, low carrier masses \cite{fan2015}, sharp absorption onsets \cite{fan2015}, long carrier lifetimes \cite{fan2015, Correa-Baena2017, chu_low-frequency_2020}, and long diffusion lengths \cite{fan2015, Xing2013, Stranks2013, Correa-Baena2017}. These favorable electronic properties are employed in optoelectronic applications, including photovoltaics \cite{fan2015, Hodes2013, Correa-Baena2017, Jeong2020}, in which power conversion efficiencies exceed 24\% \cite{Jeong2020}. However, in contrast to conventional semiconductors, electronic structure in halide perovskites is strongly coupled to large, polar, and anharmonic lattice fluctuations \cite{Egger2018,poglitsch1987, yaffe2017,whalley2016}. In room temperature MAPbI$_3$, tilting of the Pb-I octahedra and rotation of the A-site MA cation are both strongly anharmonic, creating a glass-like phonon structure with a short phonon mean free path and diffusive phonon transport \cite{poglitsch1987,Miyata2017, sharma2020}. Therefore, understanding the properties of band-like electronic states requires treating their interaction with local fluctuations in a dynamically disordered atomic lattice \cite{Egger2018,schilcher2021}. This combination of electronic and atomic structure has been linked to a wide variety of observed properties, including high defect tolerance \cite{steirer2016, Yin2014, chu_low-frequency_2020}, low radiative and non-radiative recombination leading to long carrier lifetimes and diffusion lengths \cite{Stranks2013,bi2016}, and large polaron formation \cite{zhu2015}.

The length and time scales required to simulate dynamic disorder pose a challenge to traditional electronic structure methods such as conventional density functional theory (DFT), while anharmonic fluctuations require going beyond linear electron-phonon coupling \cite{saidi_temperature_2016}. The dynamically disordered lattice~\cite{poglitsch1987} means atomic motion can be viewed in terms of local fluctuations \cite{yaffe2017}, with fluctuations of the charged lattice and bond displacements respectively leading to fluctuations in the onsite energy and hopping of localized electronic orbitals. Several previous studies have modeled halide perovskite electronic structure with tight binding methods \cite{Boyer-Richard2016,mayers2018,kim2014,ma2015,jin2012,Zheng2019,ashhab2017,Lacroix2020}, using empirical modifications of equilibrium crystal structures to calculate the effects of strain 
\cite{Boyer-Richard2016}, spin orbit coupling \cite{Boyer-Richard2016, ashhab2017}, spin splitting \cite{kim2014}, or random fluctuations in onsite energies \cite{ashhab2017,Lacroix2020} and MA orientations \cite{Zheng2019}. However, accurate prediction of finite-temperature electronic structure requires accurate prediction of the instantaneous Hamiltonian matrix elements for a realistic thermally disordered structure. 


Herein, we develop an atomic orbital tight binding model for MAPbI$_3$ which can calculate the electronic structure of systems involving 100,000s of electrons with moderate computational cost. The model predicts orbital onsite energies, hopping amplitudes, and spin-orbit couplings from atomic positions, which, in practice, are often generated from molecular dynamics trajectories. Through a tight-binding Hamiltonian, these quantities together provide a complete description of the frontier band electronic structure, which dictates carrier dynamics under normal operating conditions. This model builds upon earlier work \cite{mayers2018,McClintock2020} by adding a new set of physically motivated structural descriptors that result in increased accuracy, allowing for the investigation of electronic structure at energy scales relevant to localization and dynamic disorder in these materials. We find that a detailed description of Coulomb and short-ranged interactions is essential for an accurate description of the Hamiltonian, as is the inclusion of a number of non-bonding hopping amplitudes whose fluctuations play an important role in finite temperature electronic structure. The interpretations of these descriptors are discussed in Section~\ref{sec:model} and resulting predictions of physical observables like bandgaps and carrier masses, in Section~\ref{sec:fluc}.  

We then apply this model to large-scale simulations at finite temperatures, showing the onset of carrier localization induced by dynamic disorder. 
We study the length scale dependence of bandgap fluctuations, carrier masses, and spin splittings, finding significant disorder-induced renormalization of carrier masses and rapid reduction in spin splittings over progressively larger length scales. These results are collected in Section~\ref{sec:largen}.

\section{Descriptors of the electron-lattice interaction}\label{sec:model}

While the complexities of the electron-lattice interaction stem from the large numbers of electronic and ionic degrees of freedom, there are guiding principles that simplify the task of parameterizing these interactions. Long-range interactions arise from Coulombic forces, while short-range interactions are limited in the number of ions they can involve. In the following, we identify structural descriptors of the electron-ion interaction, and investigate the relative extents to which they control the electronic structure.

Our starting point is the tight binding Hamiltonian

\begin{equation}
\label{eq:tb}
\hat{H}_{TB} = \sum_{i} \epsilon_i(\{\vec{R}_a\})c^\dagger_i c_i + \sum_{<i,j>} t_{ij}(\{\vec{R}_a\}) c^\dagger_i c_j + h.c.   
\end{equation}

where \(\{\vec{R}_a\}\) are the classical atom positions and \(c_i\) and \(c^\dagger_i\) are the electron annihilation and creation operators for atomic orbital states. Our model consists of Pb $s$- and $p$-orbitals, and I $p$-orbitals, as these are the orbitals with dominant contributions to the conduction and valence bands. We use an atomic orbital basis rather than a maximally localized Wannier function (MLWF) basis because atomic orbitals are independent of crystal structure while MLWFs are not. This allows for easier prediction of the atomic orbital Hamiltonian from atomic positions. In our model, the orientation of $p$-orbitals is fixed to lie along the crystallographic axes. 

The purpose of the model is to predict the terms \(\epsilon_i(\{\boldsymbol{R}_a\})\) and \(t_{ij}(\{\boldsymbol{R}_a\})\) (choosing which \(t_{ij}\) are non-zero), allowing for electronic structure calculations from a set of classical atom positions. In our model, \(\epsilon_i(\{\boldsymbol{R}_a\})\) is a linear combination of potentials calculated from atomic positions and \(t_{ij}(\{\boldsymbol{R}_a\})\) is calculated from the bond geometry between the \(i\) and \(j\) atoms for bonding and non-bonding interactions, and selection rules and potentials for spin-orbit coupling. 
The numerical parameters for each matrix element are given in Supplementary Information Section 3.

\subsection{Onsite Energies}\label{sec:onsite}

As the main contribution to the onsite energies is expected to be the electrostatic potentials,
previous tight binding models have approximated onsite energy fluctuations based on the potential at the center of the atom \cite{mayers2018,Lacroix2020}. However, we find that this is insufficient because this potential does not distinguish between different $p$-orbitals on the same atom, which can have significant onsite energy variations. Furthermore, modeling onsite energies using point-charge Coulomb potentials would neglect several effects, such as electronic screening, finite orbital and ionic radii, electron-electron Hartree and exchange-correlation interactions, and hydrogen bonds, which are believed to be important in MAPbI\(_3\) \cite{lee2016}.

To address these issues, we include not only the Coulomb potential $V_C$ in the model, but also a short-range potential $V_{SR}$. We model the effective potential felt by orbitals as $V(\vec{r})=V_C(\vec{r})+V_{SR}(\vec{r})$, where 
\begin{equation}
\label{eq:vcoul}
V_C(\vec{r}) = \sum_{A} \sum_{a \in A} \frac{q_A}{4\pi\epsilon_0 |\vec{R_a} - \vec{r}|}
\end{equation}

\begin{equation}
\label{eq:vsr}
V_{SR}(\vec{r}) = \sum_{A} \sum_{a \in A} \frac{\eta_A}{8\pi l_A^3}\exp(-|\vec{r}-\vec{R}_a|/l_{A}) = \sum_{A} V_{SR,A}(\vec{r})
\end{equation}

Here, the potentials are written as a sum over the contribution from atoms of type $A$ , which belong to the set $ \{Pb, I, N, C, H(-N), H(-C) \}$.  In anticipation of the fitting done below, we define the contribution to the short-range potential from atoms of type $A$ to be $V_{SR,A}$. The charges $q_A$ and the ionic radii $l_A$ are fixed in our model (Table~\ref{tab:charges}), while the relative magnitudes of the short-range potentials $\eta_A$ are allowed to vary. The charges on the N, C, and H atoms are chosen based on their percent ionic character \cite{pauling1939} and constraining the MA dipole moment to be \(0.477 e\AA\) \cite{frost2013}. The orbital onsite energies are then given by the effective potential
\begin{equation}
\label{eq:onsite-general}
    \varepsilon_i(\{\vec{R}_a\}) = \int d^3r \, V(\vec{r}) \, \lvert \psi_i(\vec{r})\rvert^2 
\end{equation}
where $\lvert \psi_i(\vec{r})\rvert^2$ is the charge density for orbital $i$. We simplify the evaluation of Eq.~\ref{eq:onsite-general} by performing a Taylor series expansion in the spatial extent of the orbitals in the cartesian directions $\{x_1,x_2,x_3\}$, obtaining
\begin{equation}
\label{eq:tbonsite}
    \varepsilon_i(\{\vec{R}_a\}) = \sum_{n=0} \sum_{\mu=1}^3 a^{(n)}_\mu \frac{d^n V_C}{dx_\mu^n} +  \sum_{A} \sum_{n=0} \sum_{\mu=1}^3 b^{(n)}_{A,\mu} \frac{d^n V_{SR,A}}{dx_\mu^n} + \bar{\varepsilon}
\end{equation}
where the $a^{(n)}_\mu$ and $b^{(n)}_{A,\mu}$ coefficients are treated as fitting parameters, and the potentials and their derivatives are evaluated at the position of orbital $i$. We have included an explicit potential offset $\bar{\varepsilon}$ to account for the vacuum level.  The fitting parameters thus obtained can be interpreted as encoding the amount of screening of the potentials $V_C$ and $V_{SR}$, as well as the sizes of the orbital charge density $\lvert \psi_i(\vec{r})\rvert^2$ (the parameters $\eta_A$ are absorbed into $b^{(n)}_{A,\mu}$).  The spatial derivatives of $V_C$ and $V_{SR}$  capture the directional dependence of the $p$-orbital onsite energies. In our model, we include up to 4th order in $V_C$ derivatives and 2nd order in $V_{SR}$ derivatives. Symmetry requirements place restrictions on the $a^{(n)}_\mu$ and $b^{(n)}_{A,\mu}$ coefficients --- for an $s$-orbital, derivatives in all directions carry equal weight, so that $a^{(n)}_1=a^{(n)}_2=a^{(n)}_3$. For a $p$-orbital oriented along the $x_1$ axis, $a^{(n)}_1 \ne a^{(n)}_2=a^{(n)}_3$. Odd derivatives do not contribute because the orbital electron density is inversion symmetric. In our implementation, we have used $\nabla^2 V_C = 0$ to simplify the numerical evaluation of the Coulomb terms. 

By including higher order potential derivatives, the electronic structure becomes sensitive to an expanded set of structural distortions. This is of particular relevance in this material class where the soft lattice can undergo large amplitude anharmonic fluctuations. For instance, rotations of a single PbI$_6$ octahedron do not change the potential at the central Pb site, but do modify the directional potential derivatives. The ``head-to-head'' A-site modes predominant at low frequencies~\cite{yaffe2017} are likewise strongly directional in their potential fluctuations. Because of this, we find that fits lacking either directional derivatives or short ranged potentials have limited accuracy (see SI Section 5).




\subsection{Hopping Parameters}
We consider nearest-neighbor hopping parameters $t_{ij}$, which can be categorized into Pb s - I p $\sigma$-bonds, Pb p - I p $\sigma$-bonds, Pb p - I p $\pi$-bonds and Pb s - I p, Pb p - I p, and I p - Pb p nonbonding hopping parameters. The latter three nonbonding parameters are zero in an idealized perovskite crystal structure, but can take on non-zero values in a dynamically disordered crystal structure at finite temperature. We model the hopping parameters \(t_{ij}(\{\vec{R}_a\})\) with a Koster-Slater form based on the displacement $(x_1, x_2, x_3) = \vec{r}_j - \vec{r}_i$ between the two bonded atoms, with hopping parameters decreasing exponentially with bond displacement. In general, the form of the bond hopping parameters is
\begin{equation}
\label{eq:bondhop}
    t_{ij} = a \exp\left( -\sum_\mu b_\mu x_\mu -\sum_\mu c_\mu x_\mu^2\right) + d
\end{equation}
and the form of the non-bonding hopping parameters is 
\begin{equation}
\label{eq:nbhop}
    t_{ij} = a x_\nu \exp\left( -\sum_\mu b_\mu x_\mu -\sum_\mu c_\mu x_\mu^2\right)
\end{equation}
where $a$, $b_\mu$, $c_\mu$, and $d$ are fitting parameters and for non-bonding $\nu$ is the direction of the non-bonded p-orbital. This functional form captures changes in bond strength with shortening or lengthening of the bond along the primary bonding direction ($b_\mu$), as well as the effects of atom displacement along perpendicular directions ($c_\mu$). As with the onsite energies, symmetry requirements place constraints on the $b_\mu$, $c_\mu$ parameters; additionally, the parameters are used only to the leading order in each direction. For instance, a Pb s - I p $\sigma$-bond along the $x_1$ direction requires that $b_2=b_3=0$, and $c_1=0$ as it is a second order effect in the $x_1$ direction.
The functional forms for each of the six cases are given in Supplementary Information section 3.

We find that bond hopping parameter fluctuations are among the most important Hamiltonian matrix elements in finite temperature bandgap, mass, and spin splitting calculations. Additionally, we find that non-bonding hopping parameters play an important role in finite temperature fluctuations despite being zero in the time-averaged tetragonal and cubic structures with all atoms in high-symmetry positions. 

\subsection{Spin Orbit Coupling}
The spin orbit coupling (SOC) has a strong effect on the electronic structure of the halide perovskites, due to the presence of multiple high-$Z$ elements. The effects include a strong renormalization of the bandgap, possible Rashba splitting~\cite{kepenekian_rashba_2015,kepenekian2017}, circular dichroism~\cite{sercel_circular_2020}, unusual spin dynamics\cite{belykh_coherent_2019,giovanni_ultrafast_2019,wang_spin-optoelectronic_2019}, and modified excitonic fine structure\cite{becker_bright_2018}. The Rashba effect for example involves the combination of SOC with crystal structure distortions that break inversion symmetry, and has been linked to indirect bandgaps in the halide perovskites\cite{wang2017}. Spin orbit coupling fluctuations also play an important role in bandgap fluctuations. It is therefore necessary to include SOC and account for structural distortions. 

For two orbitals $\lvert\psi_i\rangle$ and $\lvert\psi_j\rangle$ in a spherically symmetric atomic potential, the SOC takes the form:
\begin{align}\label{eq:soc-general}
t_{i j} = & \left\langle \psi_i \left\lvert \frac{-e}{2m^2c^2}\frac{1}{r}\frac{dV(r)}{dr} \hat{\boldsymbol{L}} \cdot \hat{\boldsymbol{S}} \right\rvert \psi_i \right\rangle 
\end{align}
In our model, we consider SOC between onsite $p$-orbitals on Pb and I only, as $ \hat{\boldsymbol{L}} \cdot \hat{\boldsymbol{S}}$ vanishes for $s$-orbitals, and the SOC of orbitals on different sites is negligible. In a dynamically fluctuating crystalline environment, the potential in Eq.~\ref{eq:soc-general} is not fully spherically symmetric, with finite temperature fluctuations modifying each atom's central potential. We therefore modify Eq.~\ref{eq:soc-general} to
\begin{align}\label{eq:soc-tb}
t_{i j}(\{\vec{R}_a\}) = & \gamma_{ij}(\{\vec{R}_a\}) \left\langle \psi_i \left\lvert \hat{\boldsymbol{L}} \cdot \hat{\boldsymbol{S}} \right\rvert \psi_j \right\rangle 
\end{align}
where the magnitude of the SOC between orbitals on the same atom $\gamma_{ij}(\{\vec{R}_a\})$ is expected to be affected by the electrostatic potential and its derivatives in the vicinity of the atom. We therefore take it to be a linear combination of the potentials $V_C$ and $V_{SR}$ and their derivatives, as in section~\ref{sec:onsite} (although with different coefficients): 
\begin{equation}
    \gamma_{ij}(\{\vec{R}_a\}) = \sum_{n=0} \sum_{\mu=1}^3 c^{(n)}_\mu \frac{d^n V_C}{dx_\mu^n} +  \sum_{A} \sum_{n=0} \sum_{\mu=1}^3 d^{(n)}_{A,\mu} \frac{d^n V_{SR,A}}{dx_\mu^n} + \bar{\gamma}
\end{equation}
with $\bar{\gamma}$ being a constant fitting parameter. The $\hat{\boldsymbol{L}} \cdot \hat{\boldsymbol{S}}$ term in Eq.~\ref{eq:soc-tb} introduces a phase factor that is either purely real or purely imaginary, with  time-reversal symmetry imposing the following constraints on the SOC matrix elements~\cite{mayers2018}:  
\begin{align}
t_{p_x^\uparrow, p_y^\uparrow} = \, &  t_{p_x^\downarrow, p_y^\downarrow}^* \\
t_{p_x^\downarrow, p_z^\uparrow} = \, & -t_{p_x^\uparrow, p_z^\downarrow}^* \\
t_{p_y^\uparrow, p_z^\downarrow} = \, & -t_{p_y^\downarrow, p_z^\uparrow}^* 
\end{align}

\section{Methods}\label{sec:methods}


The model outlined in Section~\ref{sec:model} is parameterized from DFT calculations on crystal structures generated by classical molecular dynamics (MD). We use the molecular dynamics force field presented by Mattoni et al \cite{mattoni2015} to generate atomic trajectories in the NVT ensemble. We use experimental lattice parameters~\cite{weller2015, baikie2013}, scaled proportionately to their equilibrium MD values (calculated in cells of dimension $6\sqrt{2}\times12\times6\sqrt{2}$). For parameterization of the model, we use 40 instantaneous 2 $\times$ 2 $\times$ 2 supercell structures, taken at 1.5 picosecond intervals to minimize correlations between instances, at each of the following temperatures: 150 K, 220 K, 240 K, 260 K, 280 K, 300 K, 320 K, 340 K. The structure is orthorhombic at 150 K, tetragonal from 220 K to 320 K, and cubic at 340 K. 
DFT electronic structure is calculated along the MD trajectories, with the PBE GGA exchange-correlation functional\cite{perdew1996}, norm-conserving non-local pseudopotentials \cite{opium}, a 4 x 4 x 4 \(\Gamma\) centered k-point grid, 50 Rydberg energy cutoff, and fully-relativistic spin orbit coupling, using the Quantum Espresso\cite{Giannozzi2009} software package.
We obtain a tight-binding Hamiltonian by projecting DFT electronic structures onto the Pb s and p and I p atomic orbitals, using the Wannier90 code\cite{pizzi2020}. These Hamiltonian matrix elements are used to fit the model according to Eq.~\ref{eq:tb}.  Carrier mass and spin splittings are calculated along the \(\Gamma \rightarrow X\), \(\Gamma \rightarrow Y\), and \(\Gamma \rightarrow Z\) directions. 


\section{Thermal fluctuations of Hamiltonian matrix elements, band masses, and Rashba effects}\label{sec:fluc}

\begin{figure}[p!]
    \centering
    \includegraphics[width=0.9\textwidth]{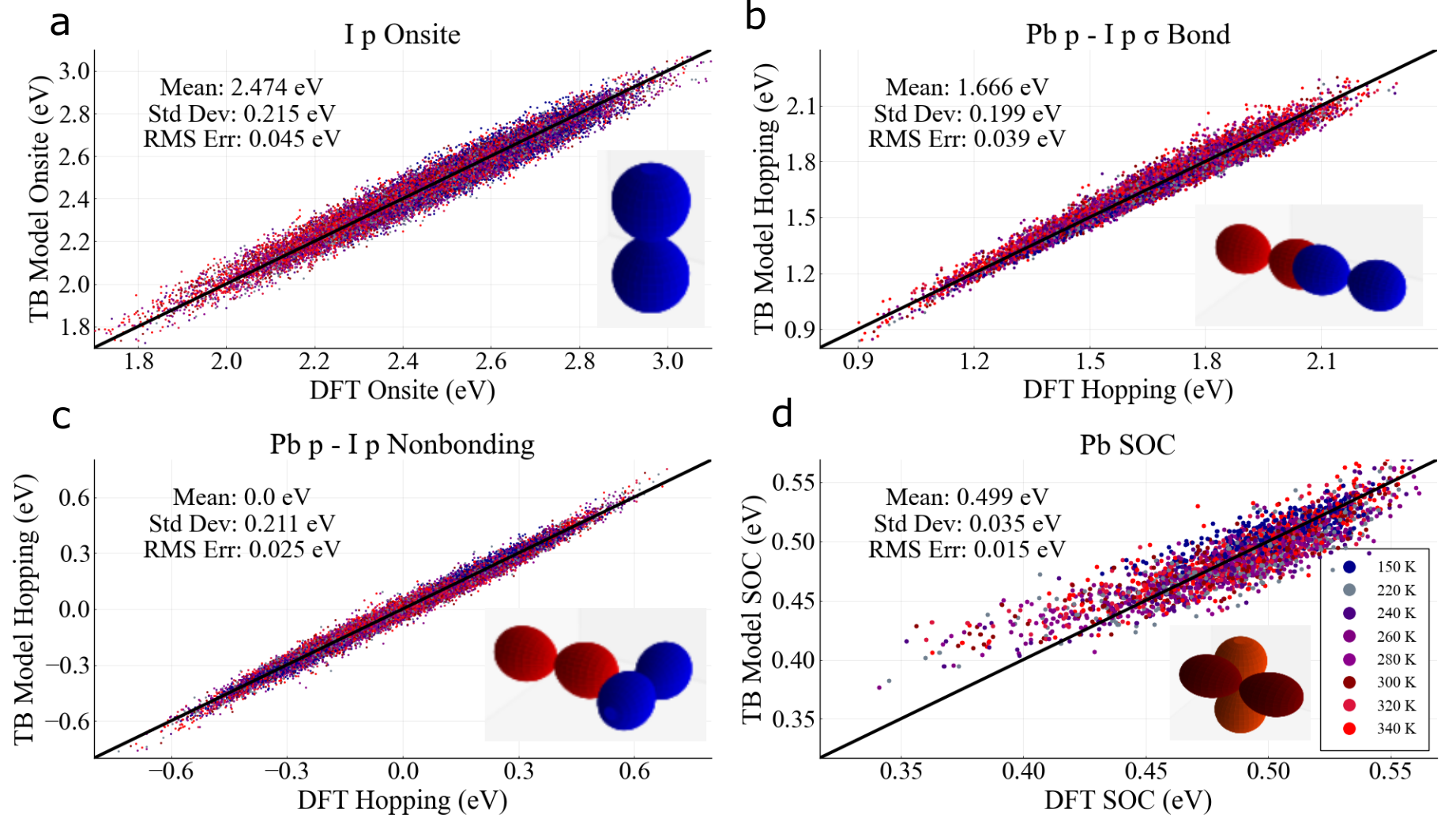}
    \caption{Scatter plots and statistics for select Hamiltonian matrix element fits in the tight binding model. Predictions of $\boldsymbol{a}$ I onsite energies, $\boldsymbol{b}$ Pbp - Ip $\sigma$ bond hopping matrix elements,  $\boldsymbol{c}$ Pbp - Ip Nonbonding hopping matrix elements, and $\boldsymbol{d}$  Pb p spin orbit coupling strength show good accuracy compared against the DFT atomic orbital Hamiltonian matrix elements.  }
    \label{fig:onsitehopstats}
\end{figure}

To understand the mechanisms by which thermal fluctuations affect the electronic structure, we compute finite-temperature carrier properties using the descriptors introduced in section \ref{sec:model}, and compare them to ab-initio predictions. Scatter plots and statistics for selected onsite energy, hopping parameter, and spin-orbit coupling fits are shown in Fig.~\ref{fig:onsitehopstats}, showing that model predictions of individual Hamiltonian matrix elements closely reproduce DFT values. Plots for the full set of matrix elements are available in SI section 4. This testifies to the high fidelity of the physically-motivated features and fits employed in our tight binding model. Furthermore, we have found that using more complex models for the tight binding parameters or using non-linear optimization improves the hopping and spin-orbit terms in the tight-binding Hamiltonian, however, these were found not to appreciably affect the accuracy of band structure predictions. 

We find that fluctuations of almost all the matrix elements along the molecular dynamics trajectories greatly exceed the thermal energy scale at room temperature, with standard deviations of 0.1-0.2 eV for the onsite energies and $\sim$0.2 eV for the bonding and non-bonding hopping amplitudes, implying that a thermally-averaged picture of the electronic structure would not be valid. Thus, off-diagonal disorder of the hopping amplitudes, including the nominally non-bonding matrix elements, cannot be neglected in this system. Even though spin-orbit coupling has the smallest fluctuations among all the matrix elements, the Pb spin-orbit matrix elements have fluctuations that are still larger than the thermal energy scale. 

We find that the model predictions of matrix elements have RMS errors in the range of 0.01-0.04 eV, depending on the matrix element. These errors are a factor of 3-10 times smaller than the fluctuations in the matrix elements arising from lattice dynamics. The I spin-orbit coupling is the sole exception, with an exceptionally small fluctuation scale of 0.005 eV, and an RMS error of a similar magnitude, which is likely limited by accuracy of the DFT calculation. Overall, these results show that the potential and bond geometry descriptors introduced above provide a faithful description of the important electronic structure fluctuations that are present at room temperature.

This picture is supported by calculations of the tight binding \(\Gamma\) point energy gap, mass, and spin splitting from the modeled matrix elements, and by comparing to predictions from the same set of matrix elements obtained from DFT (hereafter, referred to as ab initio TB). Collectively, these 3 properties describe the band structure near the band edge. We consider the band edge energy dispersion of 
\begin{equation}
E_{\pm}(\vec{k}) = \frac{\hbar^2k_{i}^2}{2m_i} \pm \alpha_i\cdot k_i
\end{equation}
which is valid for materials with time-reversal symmetry but not necessarily inversion symmetry. The Rashba splitting term \(\pm \vec{\alpha}\cdot \vec{k}\) is a spin-dependent energy splitting present when inversion symmetry is broken. We obtain the mass and Rashba \(\vec{\alpha}\) from fitting \(\frac{E_+ + E_-}{2} = \frac{\hbar^2k^2}{2m}\) and \(\frac{E_+ - E_-}{2} = \vec{\alpha}\cdot \vec{k}\), respectively, for k-points near the band edge.  

\begin{figure}[p!]
    \centering
    \includegraphics[scale=0.4]{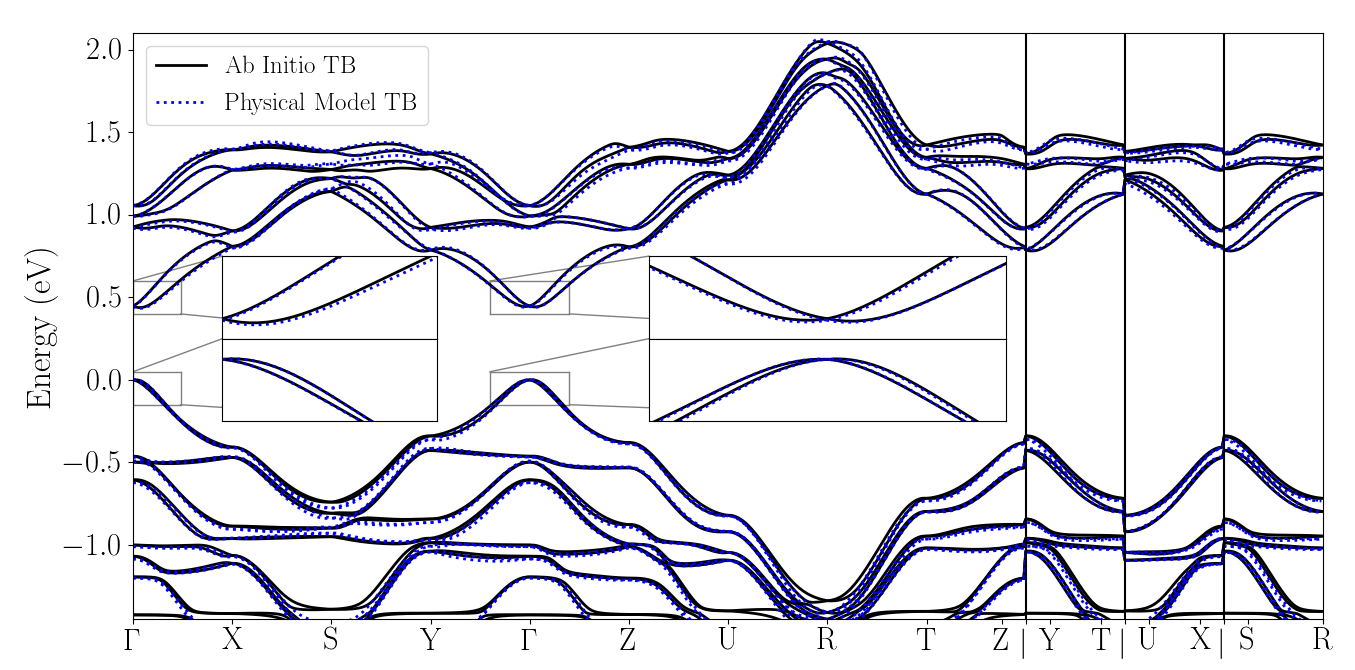}
    \caption{Band structure for an example 2 $\times$ 2 $\times$ 2 structure (the first timestep at 300 K) is shown for model TB and ab initio TB, showing that modeled tight binding matrix elements accurately reproduce results of ab initio matrix elements. The Brillouin zone vectors shown are for the 2 x 2 x 2 supercell.}
    \label{fig:examplebands}
\end{figure}

\begin{figure}[p!]
    \centering
    \includegraphics[scale=0.4]{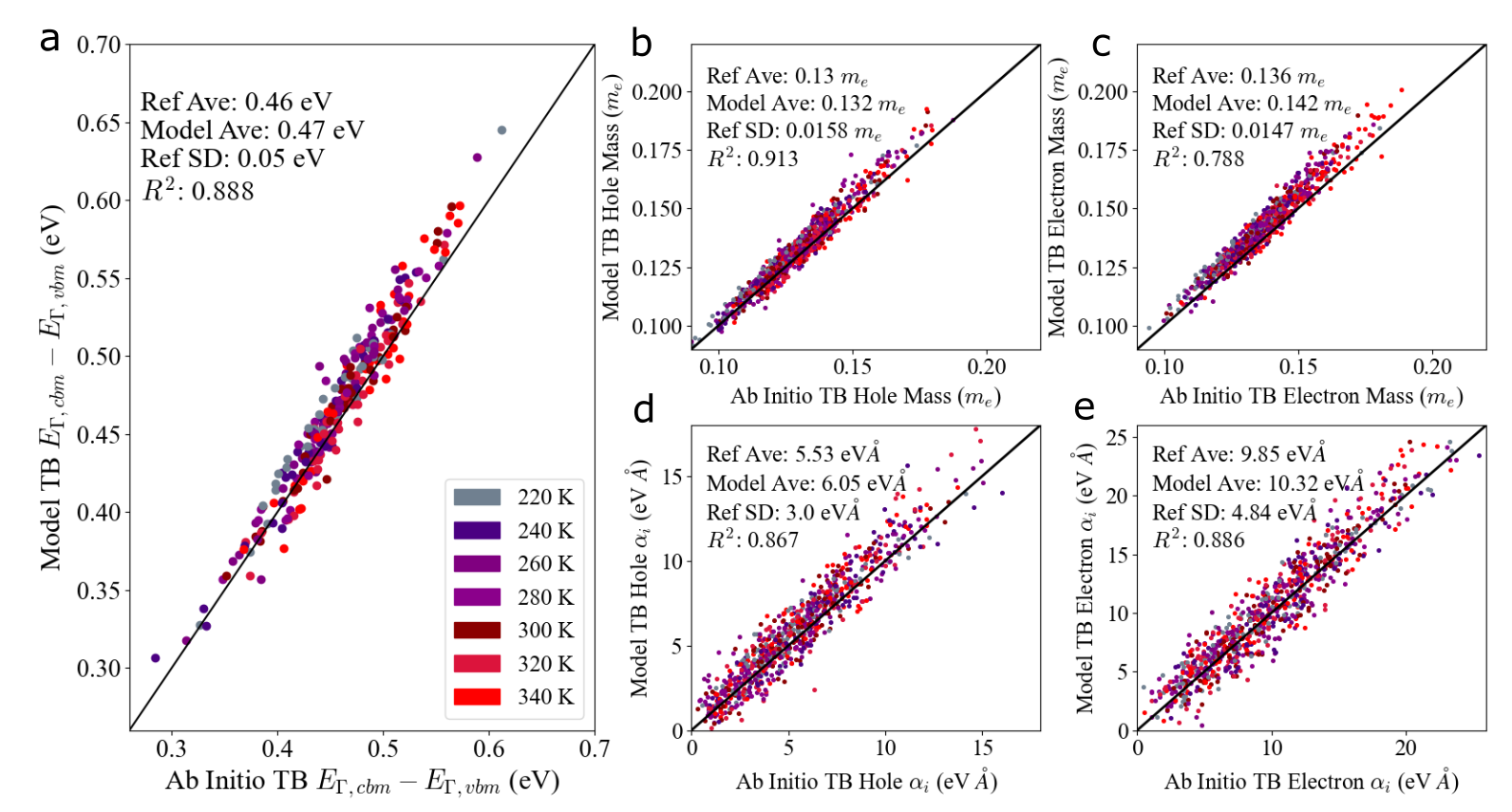}
    \caption{Tight binding band structure predictions at finite temperature,  compared to test data (tight binding performed on DFT matrix elements). a) \(\Gamma\) point energy gap, b) hole mass, c) electron mass, d) hole Rashba parameter \(\alpha_i\) , and e) electron Rashba parameter \(\alpha\). }
    \label{fig:gapmassalphapred}
\end{figure}

Since the transformation from Hamiltonian matrix elements to the band edge properties is highly nonlinear, errors in the matrix element model could in principle become magnified or give biased predictions of band edge electronic structure. For the model given above, we find that these issues are largely absent, with the predicted $\Gamma$-point bandgap, band masses, and Rashba parameters in good agreement with those calculated from ab initio TB on a $2\times2\times2$ supercell (Fig.~\ref{fig:gapmassalphapred}). 
Likewise, the band structure over the entire Brillouin zone is well reproduced by the model, as shown in Figure~\ref{fig:examplebands}  for a typical timestep along a 300 K MD trajectory. We note that the potential and geometry descriptors are not explicitly dependent on temperature, as they are functions of atomic positions only. As such, they perform well across the tested temperature range of the tetragonal and cubic phases, which they have been parameterized on. Comparisons to the full DFT band structure are given in SI section 8, where it is seen that fluctuations are generally well predicted. 

\begin{figure}[p!]
    \centering
    \includegraphics[scale=0.4]{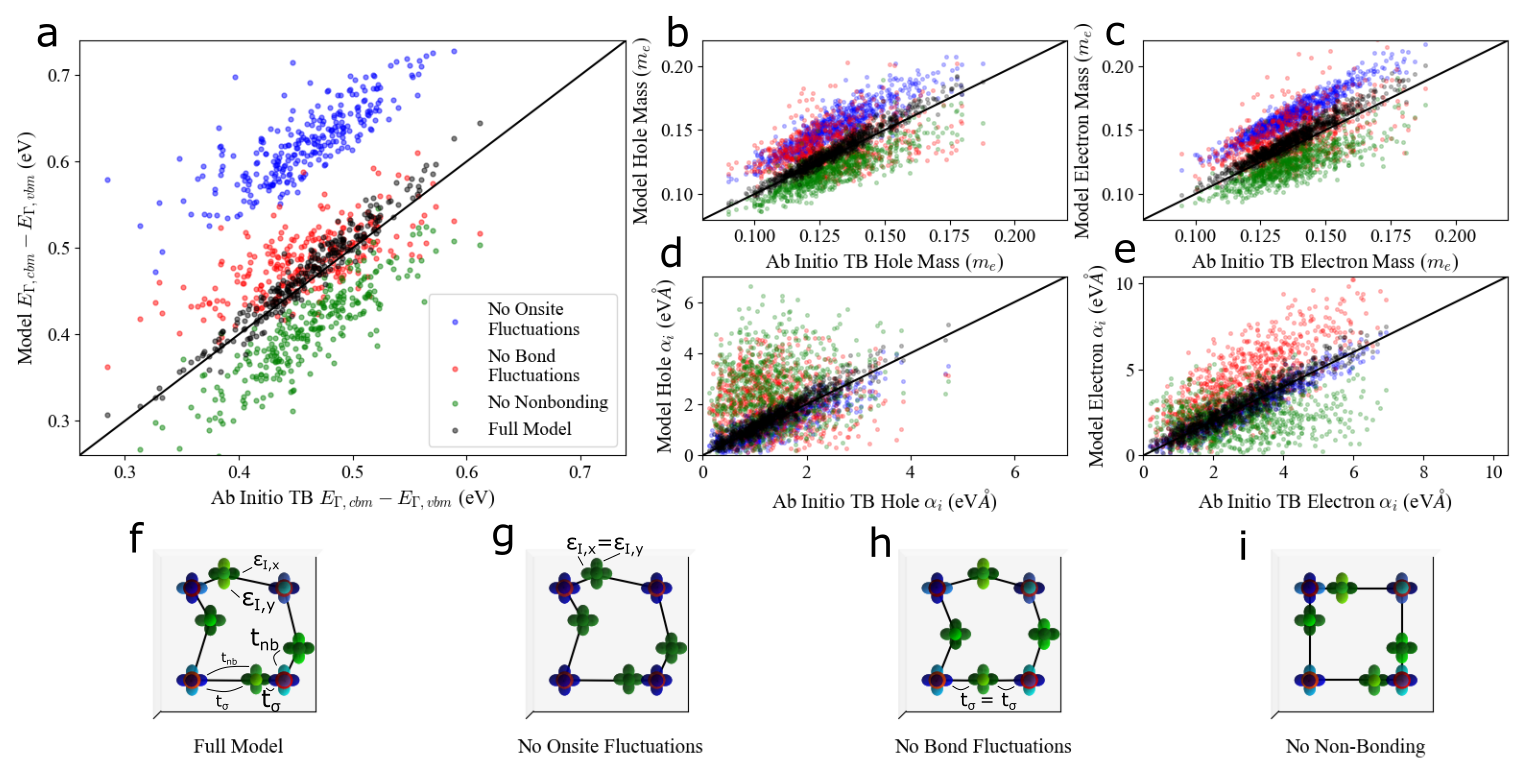}
    \caption{Effects of Hamiltonian matrix element fluctuations on band structure: (a) $\Gamma$ point energy gap, (b) hole mass, (c) electron mass, (d) hole Rashba parameter \(\alpha_i\), and (e) electron Rashba parameter. Compared to the full model, illustrated in (f), models with (g) constant onsite energies , (h) constant bond hopping, or (i) no nonbonding hopping show less accurate predictions as discussed in the text.  }
    \label{fig:alt}
\end{figure}

Next, we investigate the relative importance of the various types of fluctuations on the electronic structure, by considering a series of simplified models. By setting each onsite energy to its average value, we neglect potential fluctuations in a manner akin to a mean-field approximation. This approximation results in significantly overestimated bandgaps and band masses compared to ab initio tight-binding values on a 2 $\times$ 2 $\times$ 2 supercell (Figure~\ref{fig:alt}), implying that bandgaps and masses are strongly renormalized by the presence of potential fluctuations. This can be understood as the result of potential fluctuations increasing the bandwidth of valence conduction bands, leading to a reduced bandgap and band masses.

Instead, if non-bonding hopping amplitudes are neglected (or equivalently, assumed to not fluctuate), we find that bandgaps and band masses are both underestimated. This suggests that non-bonding fluctuations tend to decrease the bandwidths of conduction and valence bands, leading to increases in the bandgap and band masses. In contrast, bond hopping fluctuations play an important role in bandgap and mass fluctuations but do not significantly affect the average in the manner of onsite and non-bonding fluctuations (Figure~\ref{fig:alt}). Furthermore, the calculation of Rashba splitting is inaccurate if either bonding or non-bonding fluctuations are neglected, as these hopping amplitudes are crucial for describing the form of local inversion symmetry breaking~\cite{kim2014}. 

In contrast, the approximation of neglecting the fluctuations in the spin-orbit couplings (see SI section 9) does not severely affect the accuracy of Rashba parameter predictions, as these are to first order determined by the average values of the spin-orbit couplings. However, the predictions of the bandgap fluctuations are improved by the inclusion of the spin-orbit coupling fluctuations, as can be seen by comparing the quality of fit with the full model. The spin-orbit fluctuations are therefore not responsible for renormalization of the time-averaged bandgap, but do affect transient electronic structure at the sub-picosecond time scale. Overall, these results demonstrate that fluctuations in potentials, bonding, and spin-orbit coupling are mutually competing factors that renormalize band edge electronic structure, and should all be included for an accurate prediction of electronic structure.        

\begin{figure}[p!]
    \centering
    \includegraphics[scale=0.4]{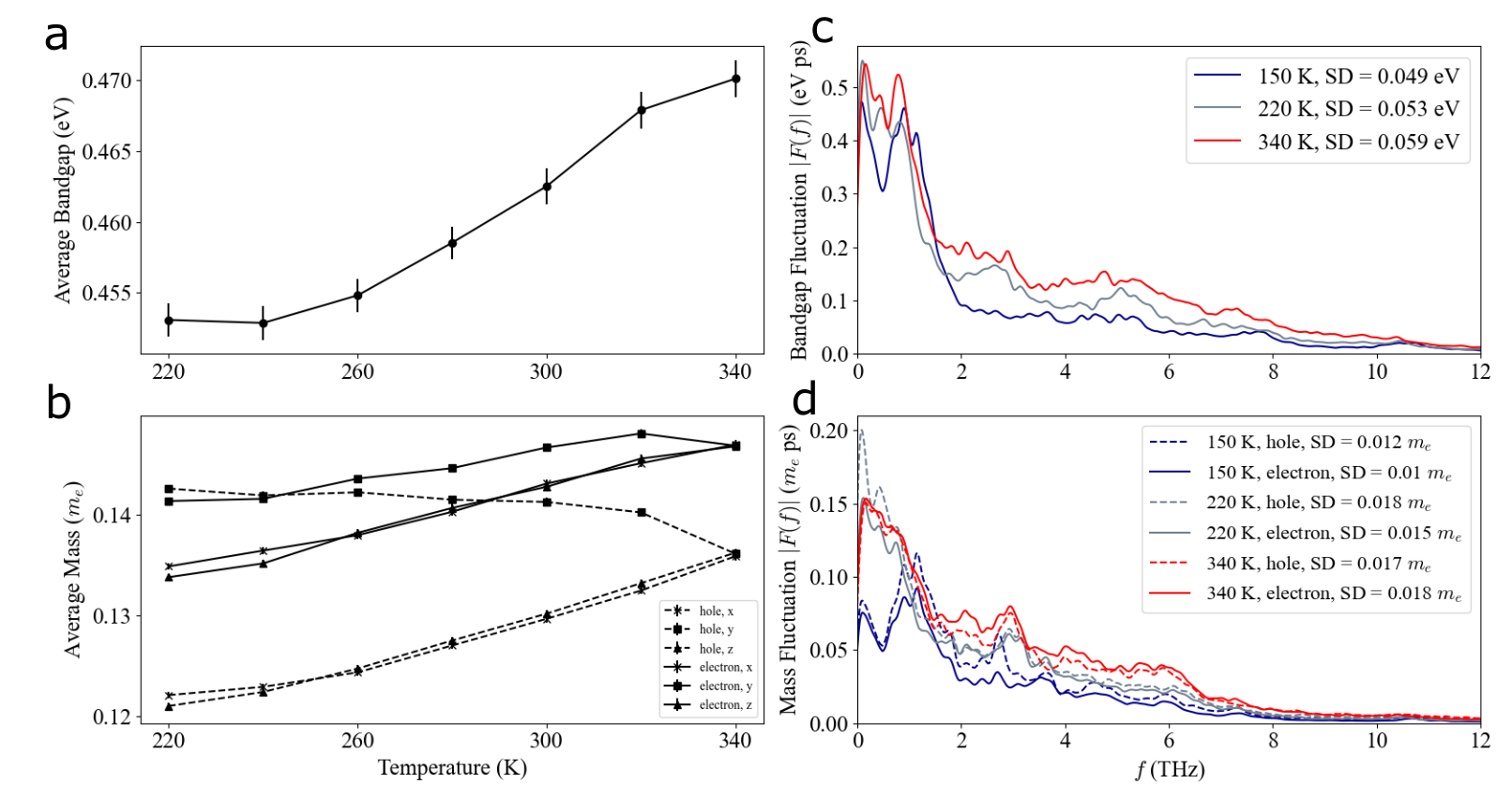}
    \caption{Temperature dependence of observables and their fluctuations. (a) and (b) Average bandgap and carrier mass as a function of temperature, calculated from a 300 ps trajectory sampled every 150 fs. Error bars show standard error. (c) and (d) Fourier transform of bandgap and carrier mass (y component) fluctuations calculated from a 300 ps trajectory sampled every 30 fs. Gaussian broaden is used with a width of 0.06 THz.}
    \label{fig:tempfourier}
\end{figure}

Lastly, we investigate the temperature dependence of bandgap and mass averages and fluctuations in 300 ps trajectories sampled every 150 fs for thermal averages and every 30 fs for fluctuation spectral analysis, a timescale accessible through tight binding. These results are shown in Fig. \ref{fig:tempfourier}. Previous work has highlighted the importance of thermal expansion and anharmonic electron-lattice interactions in predictions of the temperature dependence of the bandgap \cite{francisco-lopez2019,saidi_temperature_2016,Saidi2019,mayers2018}. Our calculations show an increase in bandgap of approximately 17 meV over a 100 K range, consistent with or somewhat below experimental temperature dependence \cite{foley2015,milot2015,francisco-lopez2019} depending on experimental method. A possible reason for this underestimate is the somewhat underestimated thermal expansion coefficient of our MD calculations compared to experiment (\(\alpha_v = 9.5*10^{-5} / \text{K}\) between 220 K to 320 K in MD compared to \(1.57*10^{-4} / \text{K}\) in Ref.  \cite{jacobsson2015}). We also find average carrier mass increases and becomes less anisotropic with increasing temperature which we attribute to a combination of disorder and thermal expansion. The bandgap standard deviation shows only a weak temperature dependence, however, higher frequency fluctuations (\(\gtrsim\) 1 THz) show a significant increase with temperature. We rationalize this result by noting that all temperatures show some degree of dynamic disorder which fluctuates at longer timescales (\(\gtrsim \) 1 ps), while increasing temperature increases the energy of short timescale fluctuations. 

\section{Effect of Dynamic Disorder on Carrier Masses}\label{sec:disordermass}

At scales smaller than the onset of localization, two major factors increase charge carrier mass at finite temperature compared to the ideal perovskite structure without thermal fluctuations. The first is the overall increase in average bond length due to thermal fluctuations which decrease the hopping between orbitals. The second is scattering caused by spatial disorder in the finite temperature lattice, with every unit cell having different bond lengths. To quantify the relative sizes of these effects, we compare the mass calculated with tight binding under 3 conditions: 1) No fluctuations, using a Hamiltonian with hopping parameters calculated from the fully symmetric ideal tetragonal perovskite structure at 300 K and averaged onsite and spin orbit coupling matrix elements, 2) spatially uniform fluctuations, obtained by averaging 300 K thermally disordered Hamiltonian matrix elements over all unit cells so that fluctuations enter only into the average of the Hamiltonian matrix elements, and 3) 300 K thermal disorder, using the fully thermally disordered 300 K structures. For these calculations, we use 100 structures taken from 300 K trajectories, at a simulation size of 8 $\times$ 8 $\times$ 8 unit cells (a scale before any significant localization occurs as shown in Sec. \ref{sec:largen}). Results are shown in Table II. In comparing the case of no fluctuations with the case of spatially uniform fluctuations, which do include the effects of octahedral rotations at an average level, we find that the increased bond length of the latter case leads to an approximately 21\% increase in electron and hole mass above the ideal tetragonal perovskite structure. Comparing to the case of full thermal disorder instead leads to an approximately 66 \% increase for holes and 71 \% increase for electrons. We therefore identify disorder as the larger factor increasing mass at finite temperature. The large effect of thermal disorder on room temperature carrier masses is consistent with previous calculations using anharmonic electron-phonon interactions \cite{Saidi2019}.

\section{Localization induced by dynamic disorder}\label{sec:largen}

Because of the prevalence of potential and hopping amplitude fluctuations, some amount of localization can be expected of electronic states in this system. 
Localization in halide perovskites has been predicted to modify finite temperature carrier properties, depending on the strength of the disorder~\cite{ashhab2017, Lacroix2020}. We proceed to quantify the extent of localization using tight binding electronic structure obtained from the DFT-based model introduced above.
We apply our tight binding model to classical MD trajectories in large supercells, calculating electronic structure on the scale of dynamic disorder, for supercells of size N = 4, 6, 8, 10, 12, 14, and 16, at temperature T=300K.

We quantify the localization of energy eigenstates at the band edge $\ket{\psi}$ with the inverse participation ratio, given in terms of the projection of $\ket{\psi}$ onto atomic orbitals $\ket{\alpha}$: 

\begin{equation}
IPR = \left(\sum_{\alpha} \lvert \bra{\alpha} \ket{\psi} \rvert^4\right)^{-1}
\end{equation}

\begin{figure}[p!]
    \centering
    \includegraphics[scale=0.40]{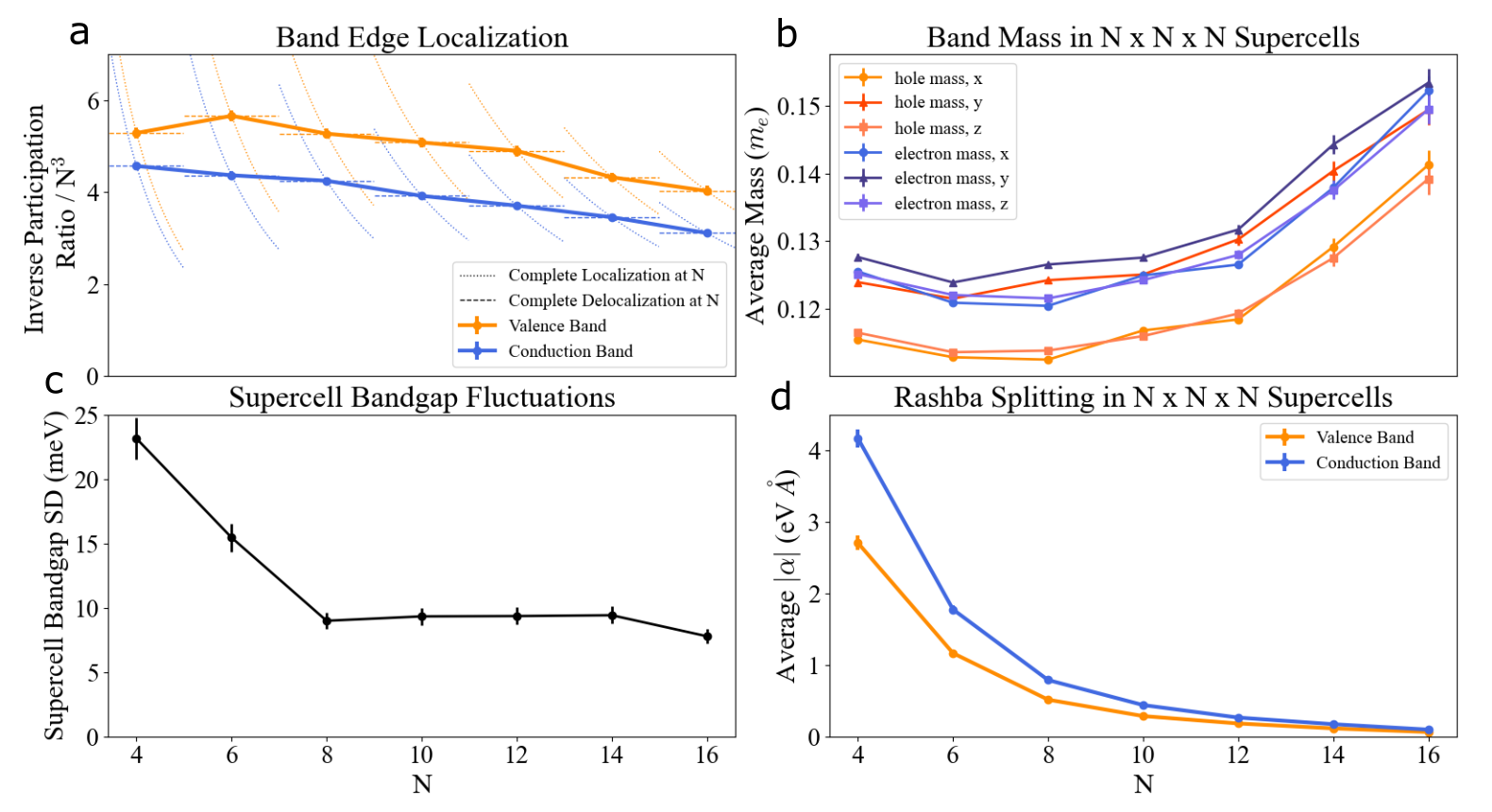}
    \caption{  (a) The average IPR / N\(^3\) decreases with supercell size but the IPR does not reach a constant, indicating only partial localization at this length scale. (b) average charge carrier mass and (c) bandgap standard deviation show effects of localization starting around n = 8, above which the mass increases and bandgap fluctuations remain relatively constant. (d) Rashba parameter magnitude as a function of supercell size, showing a quenching of spin splitting across all supercell sizes. This corresponds to a more inversion symmetric average structure. Statistics collected across 100 structures at a temperature of 300 K. Error bars for a,b,d show the standard error of the mean, while error bars in c show the standard error of the bandgap SD.}
    \label{fig:loc}
\end{figure}

\begin{figure}[p!]
    \centering
    \includegraphics[scale=0.30]{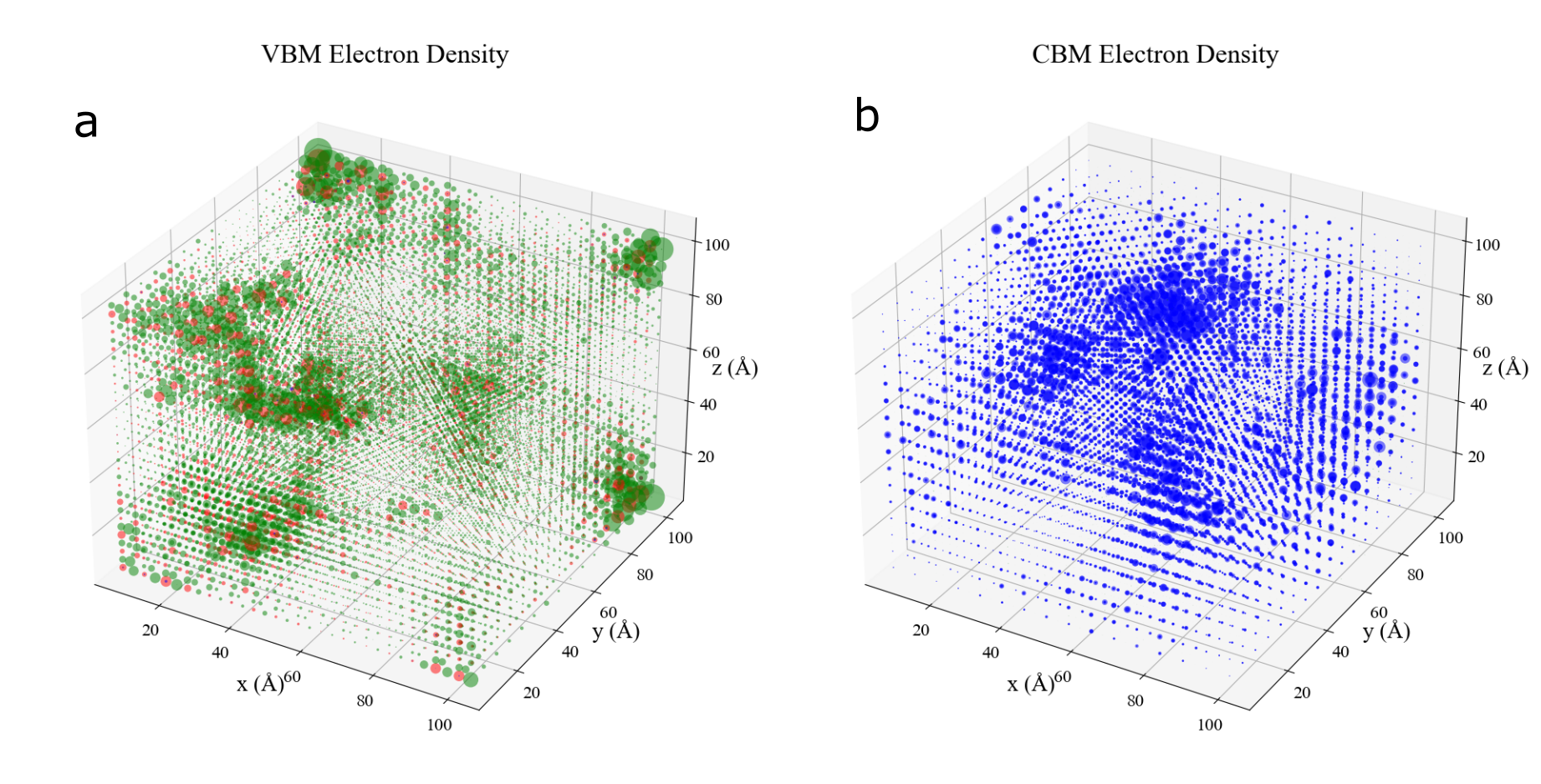}
    \caption{Plot of electron density of the (a) valence band maximum and (b) conduction band minimum of a 16 $\times$ 16 $\times$ 16 structure, simulated at a temperature of 300 K. Radius of circles is proportional to electron density of a given orbital. Red indicates Pb s, blue indicates Pb p, and green indicates I p.}
    \label{fig:densityfig}
\end{figure}

A completely localized state, i.e. one with limited spatial extent that does not change with N, will have a small IPR independent of system size. On the other hand, a completely delocalized state will have an IPR proportional to the number of atomic orbitals. Plotting the dependence of the $IPR/N^3$ on $N$ (Fig.~\ref{fig:loc}), we find that the band edge wavefunctions of halide perovskites occupy a regime intermediate between complete localization and complete delocalization. The average $IPR/N^3$ progressively decreases with $N$, indicating some amount of localization. However, this decrease is slower than the expected $\frac{1}{N^3}$ dependence for complete localization. This trend is observed for the entire range of supercells considered here, showing that the system does not reach complete localization by N = 16. Instead, the band edge wavefunctions show some spatial inhomogeneity at these length scales, but contain finite density throughout the supercells of these sizes, as illustrated in Fig.~\ref{fig:densityfig}.
The beginnings of localization behavior observed here is consistent with previous models that predict the onset of localization when disorder strengths reach a similar order of magnitude~\cite{ashhab2017,Lacroix2020}, although our predictions of the disorder strengths at 300K (potential fluctuations of 100-200 meV, Fig.~\ref{fig:onsitehopstats}) are numerically smaller than the values used in these previous models (500 meV in Ref.~\cite{Lacroix2020}). 

The increasing trend of carrier masses with supercell size further supports this picture of nascent partial localization. We find that electron and hole masses increase in all directions due to the presence of dynamic disorder, but remain finite at these supercell sizes, indicating that these states are still dispersive, and that ballistic transport is still possible at the length scales of these simulations. These results show that dynamic disorder plays a significant role in renormalizing carrier masses, with the results at $N=16$ indicating a 20\% mass increase, which is likely to increase even further at larger length scales. We find that carrier masses only start increasing significantly beyond $N=8$. This nonlinearity of the carrier mass trends suggests the presence of a length scale, which we identify as $L^*=5 nm$, corresponding to $N=8$, which marks the turning point beyond which carrier masses start to increase rapidly with supercell size.

While $L^*$ is connected to the onset of localization, the wavefunctions cannot be understood as being strictly confined to regions of size $L^*$, due to the observed IPR scaling behavior. To further understand this length scale, we investigate the scaling of the temporal fluctuations of the bandgap with system size (Fig.~\ref{fig:loc}). We find that bandgap standard deviation (SD), calculated over MD trajectories, decreases with system size at length scales less than $L^*$, but remains approximately constant for system sizes larger than $L^*$. For a system close to complete delocalization, which has almost constant wavefunction density in all unit cells, an increase in the system size can be expected to decrease the temporal fluctuations of the bandgap, due to spatial averaging across the entirety of the system. In contrast, for band edge wavefunctions that are mostly sensitive to fluctuations within a spatial region of fixed size, increasing the system size would not change the temporal bandgap fluctuations coming from dynamic disorder within that region. We therefore interpret $L^*$ as a local length scale over which band edge wavefunctions are sensitive to dynamic disorder. Even though the system contains long-range interactions (Section~\ref{sec:model}), the bandgap SD is controlled by the local fluctuations within $L^*$, while the long-range fluctuations affect its average value instead of its SD.  The results of these simulations show that halide perovskite charge carriers exhibit locality in the sense of electron-lattice interactions, while remaining dispersive and having appreciable carrier density beyond this length scale.    


Rashba splitting may play an important role in halide perovskites, caused by inversion symmetry breaking due to surfaces, defects, or thermal fluctuations \cite{frohna2018,niesner2016, kepenekian2017, mosconi2017,zheng2015,etienne2016,wang2017,hutter2017}. Bulk MAPbI\(_3\) is believed to be globally centrosymmetric, forbidding a static Rashba effect\cite{frohna2018}. However, lattice fluctuations which break inversion symmetry have been hypothesized to create a dynamical Rashba effect at a local scale \cite{etienne2016}. 
We calculate spin splitting as a function of supercell size, from the scale of local lattice fluctuations, up to 16 $\times$ 16 $\times$ 16 (10.0 nm $\times$ 10.1 nm $\times$ 10.0 nm), where the effects of average dynamic disorder rather than one specific local configuration become important. We find that spin splitting, measured by the average Rashba parameter \(|\boldsymbol{\alpha}|\), decreases approaching zero (Fig~\ref{fig:loc}d). This can be understood through the fact that the correlation length of the disorder is much smaller than $L^*$ \cite{gehrmann2019}, so that the effect of the disorder averages to be inversion symmetric. However, Rashba splitting may still play an important role near surfaces or crystalline inhomogeneities, important to thin film applications.

\section{Conclusions}
Using a physically motivated atomic orbital tight binding model, we have calculated the finite temperature electronic structure of MAPbI\(_3\) on the length and time scales of thermal disorder.
With this model, we find an increase in band gap and carrier masses with increasing temperature as well as an increase in sub-picosecond fluctuations in these observables. Investigating localization, we demonstrate that partially localized states at the band edge begin to emerge on the scale of 5-10 nm, providing a length scale for localized effects of disorder. The electronic properties of these partially localized states are controlled by dynamic disorder at the scale of $L^*= 5$ nm, even though their spatial extent is larger. Our calculations show that finite temperature effects lead to a significant increase in charge carrier mass, mostly due to disorder and also in part to increased average bond length. Thus dynamic disorder should be considered in combination with polaronic effects when considering the effects of the lattice on charge carrier mobility.  We demonstrate that spin splitting goes to zero on the scale of localization, as the correlation length of inversion symmetry breaking disorder is much less than the scale of localization. 

The relevant physics for designing a quantitatively accurate model also provides insight on the important factors determining finite-temperature electronic structure. We show that onsite energy fluctuations, bond fluctuations, and non-bonding fluctuations all play a large role in finite temperature electronic structure. The importance of non-bonding hopping parameters and onsite fluctuations driven by effects other than the Coulomb potential (including differentiating orbitals on the same atom) are especially interesting, as to our knowledge they have not been included in previous tight binding studies of dynamic disorder in halide perovskites. \cite{mayers2018,Boyer-Richard2016,Lacroix2020}.  The tight binding methods developed here are likely transferable to other materials as well as halide perovskite systems with nanoscale structure. For instance, defects \cite{steirer2016,Yin2014}, interfaces, layered structures, nanocrystals, and compounds mixing multiple halides \cite{Stranks2013,zarick2018} or cations \cite{sa2020} are all of great interest in the halide perovskite family. 

\section{Acknowledgments}
D. A. was supported by the U.S. Department of Energy, Office of Science, Office of Workforce Development for Teachers and Scientists (WDTS) under the Science Undergraduate Laboratory Internships (SULI) program. L. Z. T. was supported by the Molecular Foundry, a DOE Office of Science User Facility supported by the Office of Science of the U.S. Department of Energy under Contract No. DE-AC02-05CH11231. 
This research used resources of the National Energy Research Scientific Computing Center, a DOE Office of Science User Facility supported by the Office of Science of the U.S. Department of Energy under Contract No. DE-AC02-05CH11231.

\bibliography{apssamp}

\begin{table}[p!]
\begin{tabular}{|l|l|l| }
\hline
Atom $A$ & Charge $q_A$ (e) & Radius $l_A$ (\AA) \\ \hline
\ Pb \ & \ 2 \ & \ 1.33 \\ \hline 
\ I \ & \ -1 \ & \ 2.06 \\ \hline
\ N \ & \ 0.2509 \ & \ 0.67 \\ \hline
\ C \ & \ 0.17352 \ & \ 0.56 \\ \hline
\ H (- N) \ & \ 0.1617 \ & \ 0.53 \\ \hline
\ H (- C) \ & \ 0.03016 \ & \ 0.53 \\ \hline
\end{tabular}
\caption{Charges used to compute coulomb potentials, and atomic radii used to compute local potentials. H (- N) refers to hydrogen bonded to the nitrogen in the methylammonium ion, while H (- C) refers to hydrogen bonded to carbon.}
\label{tab:charges}
\end{table}

\begin{table}[h]
\begin{tabular}{|l|l|l|} 
\hline
\ Calculation  \ & \ Hole Mass (\(m_e\)) \ & \ Electron Mass (\(m_e\)) \\ \hline
\ No fluctuations  \ & \ (0.069, 0.074, 0.069 ) \ & \ (0.072, 0.074, 0.072) \\ \hline 
\ Spatially averaged fluctuations \ & \ (0.084, 0.090, 0.084) \ & \ (0.088, 0.089, 0.087) \\ \hline
\ 300 K Thermal Disorder  \ & \ (0.114 , 0.126, 0.115) \ & \ (0.122, 0.128, 0.123) \\ \hline
\end{tabular}
\label{tab:massdisorder}
\caption{Calculations of charge carrier mass demonstrating the effect of increased bond length and disorder at finite temperature. }
\end{table}


\end{document}